\documentclass[superscriptaddress,10pt,aps,pra,twocolumn,longbibliography, notitlepage]{revtex4-1}
\usepackage{float}
\floatstyle{boxed}
\usepackage{array}
\usepackage{graphicx}
\usepackage{amsbsy,stmaryrd}
\usepackage[utf8x]{inputenc}
\usepackage{epstopdf}
\usepackage{amsmath,amssymb,amsfonts,amsthm}
\usepackage{indentfirst}
\usepackage{dsfont}
\usepackage{soul}
\usepackage[T1]{fontenc}
\usepackage[dvipsnames]{xcolor}
\usepackage{url}
\usepackage[colorlinks]{hyperref}

\hypersetup{
	plainpages=true,
	breaklinks=true,
	hypertexnames=false,
	pageanchor=true,
	colorlinks=true,
	linkcolor={blue},
	citecolor={red},
	urlcolor={blue},
	anchorcolor={black}
}

\newcommand{\bra}[1]{\langle #1|}
\newcommand{\ket}[1]{|#1\rangle}

\newcommand{\figref}[1]{\mbox{Fig.~\ref{#1}}}

\renewcommand{\eqref}[1]{\mbox{Eq.~(\ref{#1})}}

\newcommand{\be}{\begin{equation}}
\newcommand{\ee}{\end{equation}}
\newcommand{\bea}{\begin{eqnarray}}
\newcommand{\eea}{\end{eqnarray}}

\newcommand{\LL}{\mathcal{L}}
\newcommand{\DD}{\mathcal{D}}
\newcommand{\rhot}{\hat{\rho}(t)}

\usepackage{dsfont}

\begin{document}

\title{Creating and controlling exceptional points of non-Hermitian Hamiltonians by Lindbladian invariance of homodyne detection} 

\title{Creating and controlling exceptional points of non-Hermitian Hamiltonians via homodyne Lindbladian invariance} 

\author{Fabrizio Minganti}
\email{fabrizio.minganti@gmail.com} 
\affiliation{
Institute of Physics, Ecole Polytechnique F\'ed\'erale de Lausanne (EPFL), CH-1015 Lausanne, Switzerland}

\author{Dolf Huybrechts}
\email{dolf.huyb@gmail.com} 
\affiliation{Univ Lyon, Ens de Lyon, CNRS, Laboratoire de Physique, F-69342 Lyon, France}

\author{Cyril Elouard}
\email{cyril.elouard@gmail.com} 
\affiliation{Inria, ENS Lyon, LIP, F-69342, Lyon Cedex 07, France}

\author{Franco Nori}
\affiliation{Theoretical Quantum Physics Laboratory, RIKEN Cluster for Pioneering Research, Wako-shi, Saitama 351-0198, Japan}
\affiliation{RIKEN Center for Quantum Computing (RQC), Wako-shi, Saitama 351-0198, Japan}
\affiliation{Department of Physics, University of Michigan, Ann Arbor, Michigan 48109-1040, USA}

\author{Ievgen I. Arkhipov}
\email{ievgen.arkhipov@upol.cz} \affiliation{Joint Laboratory of
Optics of Palack\'y University and Institute of Physics of CAS,
Faculty of Science, Palack\'y University, 17. listopadu 12, 771 46
Olomouc, Czech Republic}

\begin{abstract}
The Exceptional Points (EPs) of non-Hermitian Hamiltonians (NHHs) are spectral degeneracies associated with coalescing eigenvalues and eigenvectors which are associated with remarkable dynamical properties.
These EPs can be generated experimentally in open quantum systems, evolving under a Lindblad equation, by postselecting on trajectories that present no quantum jumps, such that the dynamics is ruled by a NHH. 
Interestingly, changing the way the information used for postselection is collected leads to different unravelings, i.e., different set of trajectories which average to the same Lindblad equation, but are associated with a different NHH. 
Here, we exploit this mechanism to create and control EPs solely by changing the measurement we postselect on. 
Our scheme is based on a realistic homodyne reading of the emitted leaking photons with a weak-intensity laser (a process which we call $\beta$-dyne), which we show generates a tunable NHH, that can exhibit EPs even though the system does have any in the absence of the laser.  
We consider a few illustrative examples pointing the dramatic effects that different postselections can have on the spectral features of the NHH, paving the road towards engineering of EPs in simple quantum systems.
\end{abstract}

\date{\today}

\maketitle

\section{Introduction}

The study of Exceptional Points (EPs) is at the focus of intense experimental and theoretical research~\cite{Ashida2020,El-Ganainy2018,Ozdemir2019}.
EPs are spectral singularities of non-Hermitian operators where both eigenvalues and eigenvectors coalesce~\cite{KatoBOOK}, thus characterizing the dynamics of open quantum systems \cite{Minganti2019}.
Such a singularity is mathematically associated with the nontrivial topological structure of the eigenvalue manifold, and the EP corresponds to a branching point of the solution of the characteristic polynomial of the corresponding non-Hermitian operator.

Beyond their theoretical and mathematical interest, EP's interest among physicists sparked from the discovery of parity-time ($\cal PT$) symmetry-breaking, leading to the characterization of $\cal PT$ non-Hermitian Hamiltonians (NHHs)~\cite{Bender1998} and to the study of phase transitions in finite-dimensional systems~\cite{Bender1998,Heiss1991}. 
Many experiments confirmed and demonstrated the unique properties of EPs and their influence on system dynamics, e.g.,
unidirectional invisibility~\cite{Lin2011,Regen2012},  lasers with
and enhanced-mode selectivity~\cite{Feng2014,Hodaei2014},
low-power nonreciprocal light transmission
\cite{Peng2014,Chang2014}, thresholdless phonon lasers
\cite{Jing2014,Lu2017}, enhanced light-matter interactions
\cite{Liu2016,Chen2017,Hoda2017},  loss-induced lasing
\cite{Brands2014a,Peng2014a}. The nontrivial properties of the EPs have also been studied and analyzed in
electronics~\cite{Schindler2011}, optomechanics
\cite{Jing2014,Harris2016,Jing2017}, acoustics
\cite{Zhu2014,Alu2015}, plasmonics~\cite{Benisty2011}, and
metamaterials~\cite{Kang2013}.  
At an EP, systems are also known for exhibiting nontrivial topological and localization properties, particularly in 1D and higher-dimensional  lattice architectures~\cite{LeykamPRL17,GonzalesPRB17,HuPRB17,GaoPRL18,LiuPRL19,Zhou2018,BliokhNat19,MoosSciPost19,Ge2019,Yoshida2019,Ashida2020,Gong2018,Kawabata2019,Mcdonald2021non,Nie2021,Bergholtz2021,arkhipov2022a}.
For extensive reviews see, e.g., Refs.~\cite{Ozdemir2019,Miri2019,Ashida2020,Bergholtz2021} and references therein.

Many of the previously-cited works dealt with ``semiclassical'' configurations, where the equation of motion of a strong coherent field can be mapped onto effective Schr\"odinger equations, leading to the appearance of Hamiltonian EPs (HEPs)~\cite{Ozdemir2019,Liang2017,Ashida2020,Miri2019}. 

When fully taking into account \textit{quantum dissipative processes}, it is necessary to include the action of quantum jumps (Langevin noise), often significantly changing the dynamics of a quantum system \cite{Lau2018,Langbein2018,Zhang2019,Minganti2019,Minganti2020}.
To circumvent such a problem, and witness the EPs of a quantum NHH, two experimental strategies have been recently realized: dilation of a non-Hermitian Hamiltonian in a larger Hermitian space \cite{Teimourpour2014,Yang2020} and  post-selection \cite{Naghiloo19,Chen2021q}. 
In the later case, EPs can be seen as one of the manifestations of the non-trivial dynamics of quantum systems under post-selection \cite{Kwiat95,Campagne-Ibarcq14}.

Indeed, under quite general hypotheses, the dynamics of open quantum systems can be described by a Lindblad master equation that, in turn, can be separated into the action of an effective NHH, periodically interrupted by abrupt events called quantum jumps \cite{BreuerBookOpen,walls_milburn_2011}.
According to this quantum trajectory picture, the Lindblad master equation is simply the description of the average dynamics of a system continuously probed by a set of detectors (modelling the environment), each one associated with a jump operator \cite{Molmer93,Dalibard92,HarocheBook}.
By postselecting those trajectories where no quantum jump occurs (no detector clicks), the spectral properties of the NHH can be investigated \cite{Naghiloo19}.
Within this postselected approach, the presence of EPs admits thus a simple and fascinating explanation: the very information gained by the fact that a quantum jump did not occur induces a nonunitary state update that, in turn, introduces the non-Hermiticity necessary to witness the wanted EP.

The formulation of quantum jumps, however, is not unique, 
and several 
quantum trajectory equations can be associated with the same Lindblad master equation \cite{HarocheBook,walls_milburn_2011,BreuerBookOpen}. 
Even if, usually, the quantum jumps and the NHH are represented in a ``standard'' form -- the jump operators are chosen to be orthonormal and traceless -- there exist a whole class of transformations, changing the effective Hamiltonian and the jump operators, which recover the same Lindblad master equation.
However, the dynamics at a single trajectory level can drastically change according to the form of the effective Hamiltonian and of the jump operators stemming from these transformations \cite{BartoloEPJST17,RotaNJP18,MunozPRA19}.
The different forms of quantum trajectories, associated with a given Lindblad master equation, admit a clear physical interpretation and are called unraveling: they correspond to different ways to collect the information leaking from the system into the environment. 
Although the unconditional evolution (averaged overmthe detectors' output) is unchanged, these different monitorings modify the way the system behaves along single runs of an experiment, which are conditioned on a given sequence of detector outputs.
This striking effect was experimentally demonstrated in, e.g., Ref.~\cite{Campagne-IbarcqPRX16}.

Since monitoring the occurence of quantum jumps is the key ingredient for postselection, and the way the jumps operators act modifies the non-Hermitian Hamiltonian, a natural question is what effects can be witnessed with these different unravelings of the same dissipative dynamics.
In this article, we demonstrate that different NHHs associated with the \textit{same Lindlbad} master equation display \textit{completely different} spectral properties.
In particular, by modifying the form of the quantum jumps (i.e., the way quantum information is collected) and postselecting those trajectories where no quantum jump occurred, we can induce an EP or modify its properties.

The article is structured as follows.
In Sec.~\ref{Sec:Invariances} we introduce the Lindblad master equation and its invariances, leading to the different quantum trajectories associated with the same dynamics, as well as their postselection.
We then introduce our first example in Sec.~\ref{Sec:ExampleI}, where we consider a two-level system (qubit) with gain and losses that does not display any EPs using the ``standard'' representation of the quantum jumps.
However, tuning the previously-introduced canonical transformations of the Lindblad master equation one can induce an EP.
We then show in Sec.~\ref{Sec:ExampleII} that it is possible to induce an EP in a driven Kerr resonator with a similar procedure, but just in the presence of photon loss.
Finally, in Sec.~\ref{Sec:ExampleIII}, we show that the canonical transformation can also be used to tune the properties of an EPs.
We present our conclusions in Sec.~\ref{Sec:Conclusions}.

\section{Lindblad invariances and quantum trajectories}
\label{Sec:Invariances}

The state of an open quantum system is captured by its density matrix $\rhot$. If the system interacts with a Markovian (memoryless) environment, and within the Born approximation, $\rhot$ evolves under the Lindblad master equation, which reads \cite{BreuerBookOpen,walls_milburn_2011}
\begin{equation}
    \frac{\partial \rhot}{\partial t} = \LL \rhot = -i \left[\hat{H}, \rhot \right] + \sum_{\mu} \gamma_\mu \DD \left[\hat{J}_\mu\right] \rhot .\label{Eq.:Lindblad}
\end{equation}
In this description, $\hat{H}$ is an Hermitian operator describing the coherent evolution of the system, while $\hat{J}_\mu$ are the jump operators describing the dissipation induced by the environment via the action of the dissipators, which reads
\begin{equation}
    \DD \left[\hat{J}_\mu\right] \rhot =  \hat{J}_\mu \rhot \hat{J}_\mu^\dagger - \frac{\hat{J}_\mu^\dagger \hat{J}_\mu \rhot + \rhot \hat{J}_\mu^\dagger \hat{J}_\mu}{2}.
\end{equation}

\subsection{Quantum trajectories}
\label{Subsec:Postselection}
From a theoretical point of view, the Lindblad master equation is a particular form of a quantum map \cite{WisemanBook}.
On a general ground, any quantum map can be rewritten in terms of its Kraus operators, and
\begin{equation}
    \hat{\rho}(t+dt)  = \sum_{\nu} \hat{K}_{\nu} \rhot \hat{K}_{\nu}^\dagger,\label{Eq:Kraus}
\end{equation}
where the condition
\begin{equation}\label{Eq:condition_Kraus}
     \sum_{\nu} \hat{K}_{\nu}^\dagger \hat{K}_{\nu} = \hat{\mathds{1}}
\end{equation}
is required to ensure that the quantum map is CPTP (i.e., Completely Postitive and Trace Preserving).
Given the form of \eqref{Eq.:Lindblad}, one can verify that a set of Kraus operators that recover the Lindblad master equation is
\begin{equation}
\begin{split}
    \hat{K}_{0} &= \hat{\mathds{1}} - i \hat{H} dt - \sum_{\mu} \gamma_\mu \frac{\hat{J}_\mu^\dagger \hat{J}_\mu}{2}  \, dt = \hat{\mathds{1}} - i \hat{H}_{\rm eff} dt,  \\
    \hat{K}_{\nu} &= \sqrt{\gamma_\mu dt} \hat{J}_\nu,
\end{split}    \label{Eq.:KrausOp}
\end{equation}
where the non-Hermitian Hamiltonian (NHH) $\hat{H}_{\rm eff}$ is
\begin{equation}\label{Eq:NHH}
    \hat{H}_{\rm eff} = \hat{H} - i \sum_\mu \gamma_\mu \frac{\hat{J}_\mu^\dagger \hat{J}_\mu}{2}.
\end{equation}
Each Kraus operator can be associated with one of the possible outcomes of a measurement process, whose backaction on the system is associated with the $\hat{K}_{\nu}$ operator.
In this regard, the Lindblad master equation can be interpreted as the dynamics of a system upon the continuous action of several measurement instruments.
Whenever the outcome $\nu\neq 0$ is obtained at time $t$, one of the detectors ``clicks'' and the system evolves under the action of $\hat{K}_\nu$. If none of the detectors click, the system evolves according to $\hat{K}_{0}$, as the absence of clicks still conveys some information about the system state \cite{HarocheBook}, yielding \eqref{Eq:Kraus} when the average over the measurement outcomes is taken.

From this interpretation it is natural to introduce quantum trajectories.
By keeping track of the measurement outcomes (the sequence of quantum jumps), we can exactly reconstruct the state of a given system initialized in a pure state $\ket{\Psi(t=0)}$~\cite{WisemanBook}.
If the $\hat{K}_0$ acts, the time (unnormalized) evolution of the system is described by
\begin{equation}
    \partial_t \ket{\Psi(t)} =  - i \hat{H}_{\rm eff} \ket{\Psi(t)},
\end{equation}
while, if the $\nu$th Kraus operator acts, the system evolves as
\begin{equation}
    \ket{\Psi(t+dt)} \propto \hat{J}_\nu \ket{\Psi(t)}. 
\end{equation}

If we consider the dynamics where no quantum jumps occurs, the system evolves solely under the action of the NHH $\hat{H}_{\rm eff}$.  
Thus, a postselection on those trajectories with no quantum jumps reveals the spectral properties of the NHH $\hat{H}_{\rm eff}$ in \eqref{Eq:NHH}.
In particular, we can define the eigenvectors $\ket{\Psi_j}$ and the associated ``eigenenergies'' $E_j$ such that 
\begin{equation}
    \hat{H}_{\rm eff} \ket{\Psi_j} = E_j \ket{\Psi_j}.
\end{equation}
An EP of the NHH is then defined as a point where $E_j = E_k$ and $\ket{\Psi_j} = \ket{\Psi_k}$.
Higher-order degeneracies can take place, and for this reason one calls the order of an EP the number of coalescing eigenvectors (e.g., $\ket{\Psi_j} = \ket{\Psi_k}=\ket{\Psi_l}$ is an EP of order 3).

As it has been experimentally shown in Ref.~\cite{Naghiloo19},  this postselection procedure allows studying the emergence of the \textit{Hamiltonian EPs}, i.e., the degeneracy of the NHH, by reconstructing the evolution of the quantum system.
In principle, one needs perfect detectors that detect the jumps with unitary efficiency and exhibit no dark counts.
In the presence of finite-efficiency detectors, however, one can still observe the effects of the EP of NHH by analyzing the associated hybrid-Liouvillian \cite{Minganti2020}.

Notice that, just as for any quantum dynamics, in order to experimentally assess the properties of a system evolving at an EP, it is not sufficient to just know that no jump occurred.
Indeed, the simple measurement of the jump operators does not not yield all the information about the system state, allowing  to completely characterize the properties of the system.
For instance, in a qubit system, determining the presence of an EP amounts to (i) postselect the trajectory where no jump happened, and (ii) perform a measurement of the system at a given time $t$; (iii) Repeat the measurement for different runs (having initialized the system in the same state); (iv) repeat the same procedure for several different times $t$.
This procedure highlights the presence of an anomalous dynamics associated with an EP.

\subsection{Lindbladian invariances, measurements, and new trajectories}

Jump operators are usually chosen to be traceless ($\operatorname{Tr}[\hat{J}_{\mu}] = 0$) and orthonormal ($\operatorname{Tr}[\hat{J}_{\mu} \hat{J}_{\nu}] \propto \delta_{\mu \nu}$). 
Such a choice, although mathematically convenient, should not be privileged from a physical point of view. Indeed, the set of jump operators, associated with a given Lindblad equation, is not uniquely determined.

Consider, for instance, the affine transformations
\begin{equation}\label{Eq:Lindblad_invariance_2}
\left\lbrace
\begin{split}
    \hat{J}_{\mu}' &= \hat{J}_{\mu} + \beta_\mu \hat{\mathds{1}}, \\
    \hat{H}' &= \hat{H}  - \sum_{\mu} \frac{i \gamma_\mu}{2} \left( \beta_\mu^* \hat{J}_\mu - \beta_\mu \hat{J}_\mu^\dagger \right).
\end{split} 
\right.
\end{equation}
Although \eqref{Eq:Lindblad_invariance_2} modifies both the Hamiltonian and the jump operators, it does not change the Lindblad master equation result.
Indeed, one can easy show that, the Lindblad master equation stemming from $\hat{H}'$ and $\hat{J}_{\mu}'$ is the same as \eqref{Eq.:Lindblad} \cite{BreuerBookOpen}.
However, for $\beta\neq 0$, the effective Hamiltonian changes as
\begin{equation}\label{Eq:NHH_beta}
\begin{split}
\hat{H}_{\rm eff}(\beta) &= \hat{H}  - \sum_\mu \frac{i \gamma_\mu}{2} \left( \beta_{\mu}^* \hat{J}_\mu - \beta_{\mu} \hat{J}_\mu^\dagger \right) \\ & \quad  -  \sum_\mu\frac{i \gamma_\mu}{2} \left( \hat{J}_\mu + \beta_{\mu} \hat{\mathds{1}} \right)^\dagger \left( \hat{J}_\mu + \beta_{\mu} \hat{\mathds{1}} \right) .
\end{split}    
\end{equation}

In quantum optics, the quantum trajectory stemming from the orthonormal set $\hat J_\nu \equiv \hat{a}(\omega)$ [where $\hat{a}(\omega)$ is the annihilation operator of the mode at frequency $\omega$] describes a situation in which the photons emitted by the system are instantaneously detected by a photon-counter, while the set of jumps $\hat J'_\nu$, defined by \eqref{Eq:Lindblad_invariance_2}, is relevant in the case of a homodyne detection setup. 
In the latter case, the emitted photons are mixed with a coherent field on a beam-splitter before being detected. 
For
usual homodyne detection, one chooses $\vert\beta\vert^2\gg 1$, while here we will consider finite values of $\beta$ (associated with a weaker coherent field). 
For this reason we call such a trajectory resulting from \eqref{Eq:Lindblad_invariance_2} and from  \eqref{Eq:NHH_beta} a  $\beta$-\textit{dyne} unraveling (see also Appendices~\ref{App:Homodyne}~and~\ref{App:BetaDyneModel} for a theoretical model of this detection setup, its physical interpretation, and the derivation of the associated dynamics).

It is rather remarkable that the affine transformation in \eqref{Eq:Lindblad_invariance_2} produces an ambiguity in the realization of the system's quantum trajectories.
For instance, consider again a bosonic system characterized, this time, by just one annihilation operator $\hat J_\nu = \hat{a}$ {(i.e., the environment induces dissipation in the form of particle losses described by $\DD [\hat{a}]$ \textit{in the average dynamics}).} 
{The effect of the betadyne unraveling at a \emph{single trajectory} level is that (i) the action of the jumps operator {$\hat J'_\nu$} do not eject one full photon from the system and (ii) when the jump {$\hat J'_\nu$} does not occur the bosonic field within the cavity gets displaced by a coherent amplitude $\gamma_\mu \beta /2$, according to \eqref{Eq:Lindblad_invariance_2}}.
What are then the effects of this invariance {on the spectral properties of the non-Hermitian Hamiltonian}, and can they be detected experimentally? 
That is, can one practically implement a postselection procedure that, even if the average dynamics is purely incoherent, exhibits a significantly different dynamics at the NHH level?
Below, we show that the transformation \eqref{Eq:Lindblad_invariance_2} can profoundly change the structure of the NHH, even inducing the presence of EPs in otherwise non-degenerate Hamiltonians.
Owing to the postselected measurement, we prepared the system at an EP, and the ``standard'' measurement protocol on the system, i.e., the points (i)-(iv) described in Sec.~\ref{Subsec:Postselection}, can demonstrate the properties of the system.

Even if in the following, we will focus on the $\beta$-dyne transformation induced by \eqref{Eq:Lindblad_invariance_2}, for the sake of completeness let us notice that not all transformations on the Lindblad master equation modify the structure of the NHH.
For instance, the unitary transformation
\begin{equation} \label{Eq:Lindblad_invariance_1}
    \sqrt{\gamma'_{\mu} }\hat{J}_{\mu}' = \sum_{\nu} R_{\mu, \nu} \sqrt{\gamma_{\nu} } \hat{J}_{\nu} 
\end{equation}
leaves both the Lindlbad master equation and the NHH unchanged if $R_{\mu, \nu}$ is a unitary matrix.
Indeed,
\begin{equation}\begin{split}
   \hat H'_{\rm eff} &= \hat H-\sum\limits_\mu \frac{i \gamma'_\mu}{2} \hat J'^{\dagger}_\mu\hat J'_\mu \\
    & =\hat H- \sum\limits_{\mu,\nu,\chi} R^{*}_{\mu,\nu}R_{\mu,\chi} \frac{i \sqrt{\gamma_\nu\gamma_\chi}}{2}\hat J^{\dagger}_\nu\hat J_\chi   \\ & = \hat H -\frac{i}{2}\sum\limits_{\mu}\gamma_\mu\hat J^{\dagger}_\mu\hat J_\mu = \hat H_{\rm eff},
    \end{split}
\end{equation}
because $R_{\mu,\nu}$ is unitary, i.e., $\sum\limits_\mu R^*_{\mu,\nu}R_{\mu,\chi}=\delta_{\nu,\chi}$.
Thus, if we ``mix'' the leaking fields, all postselected dynamics result in the same NHH.

\section{Example I: inducing an EP by postselection of a $\beta$-dyne trajectory}
\label{Sec:ExampleI}

As a starting point for our discussion, let us consider a two level system, whose most general NHH reads
\begin{equation}
    \left(
\begin{array}{cc}
 a & b \\
 c & d \\
\end{array}
\right) =     \left(
\begin{array}{cc}
 \tilde{a} & b \\
 c & 0 \\
\end{array}
\right) +d     \left(
\begin{array}{cc}
 1 & 0 \\
 0 & 1 \\
\end{array}
\right),
\end{equation}
where $\tilde{a} = a -d$.
The eigenvalues read
\begin{equation}
    \frac{2d + \tilde{a}\pm\sqrt{\tilde{a}^2+4 b
   c}}{2},
\end{equation}
and the (unnormalized) eigenvectors are
\begin{equation}
\left\{
 \tilde{a} \pm \sqrt{\tilde{a}^2+4 b c} , \, 2
   c 
\right\}.
\end{equation}
Consequently, the condition to observe an EP in a two-level system can be recast as
\begin{equation}
    \tilde{a} = 2 i \sqrt{b c}.
\end{equation}

\subsection{EP of a qubit with loss and gain}

Consider a two level system whose Hamiltonian is
\begin{equation}
    \hat{H} = \frac{\omega}{2} \hat{\sigma}_z,
\end{equation}
 and with jump operators
\begin{equation}
    \hat{J}_1 = \sqrt{\gamma}_-\hat{\sigma}_-,\quad \hat{J}_2 = \sqrt{\gamma}_+\hat{\sigma}_+,
\end{equation}
where $\hat{\sigma}_z$ is the $z$ Pauli matrix and $\hat{\sigma}_{\pm}$ are the raising and lowering qubit operators, respectively.
If we monitor the jump operator $\hat{J}_1$, then the effective Hamiltonian of this system is
\begin{equation}
\hat{H}_{\rm eff} = \frac{1}{2} \left(
\begin{array}{cc}
  \omega -i \gamma_- & 0 \\
 0 & - \omega - i \gamma_+ \\
\end{array}
\right).
\end{equation}
Consequently, this model can \textit{never} display any exceptional point, because the eigenvalues are always distinct and separated.

Let us now consider the $\beta$-dyne detection of both the jump operators with the same intensity $\beta$, via the transformation
\begin{equation}
    \hat{J}_1' = \sqrt{\gamma}_- \left(  \hat{\sigma}_- + \beta \hat{\mathds{1}} \right),\quad \hat{J}_2' = \sqrt{\gamma}_+ \left(  \hat{\sigma}_+ + \beta \hat{\mathds{1}} \right).
\end{equation}
The corresponding effective Hamiltonian is

\begin{equation} \begin{split}\label{Eq:Heff_beta_1}
    \hat{H}_{\rm eff}(\beta) =& - i \frac{|\beta|^2  (\gamma_- + \gamma_+) }{2} \hat{\mathds{1}}  \\ &  + \frac{1}{2} \left(
\begin{array}{cc}
\omega - i \gamma_-  & - 2 i \beta ^* \gamma_+  \\
 - 2 i \beta ^* \gamma_- & - \omega - i   \gamma_+  \\
\end{array}
\right).
\end{split}
\end{equation}
The NHH $\hat{H}_{\rm eff}$ has now the right structure to \textit{display an EP}.
In particular, the eigenvalues and eigenvectors
read
\begin{widetext}
\begin{equation}\label{Eq:Values_1}
\begin{split}
    E_{1, 2} =& -i \frac{\gamma _-+\gamma _+ +2 |\beta|^2  \left(\gamma _-+\gamma _+\right)  \pm  \sqrt{16 \gamma _- \gamma _+
   \left(\beta ^*\right)^2+\left(\gamma _--\gamma _++2 i \omega
   \right){}^2}}{4}.
    \end{split}
\end{equation}
\begin{equation}\label{Eq:Vectors_1}
\begin{split}
    \ket{\Psi_{1,2}} = \left \{ 2 i \omega \gamma _--\gamma _+  \pm \sqrt{16 \gamma _- \gamma _+
   \left(\beta ^*\right)^2+\left(\gamma _--\gamma _++2 i \omega
   \right){}^2} , 4 \gamma _- \beta ^*  \right\}
   \end{split}
\end{equation}
\end{widetext}
The EP emerges when 
\begin{equation}
    \beta =\pm i \frac{\gamma
   _--\gamma _+ - 2 i \omega}{4 \sqrt{\gamma _- \gamma _+}}. 
\end{equation}

\begin{figure}[t]
    \centering
    \includegraphics[width=0.49 \textwidth]{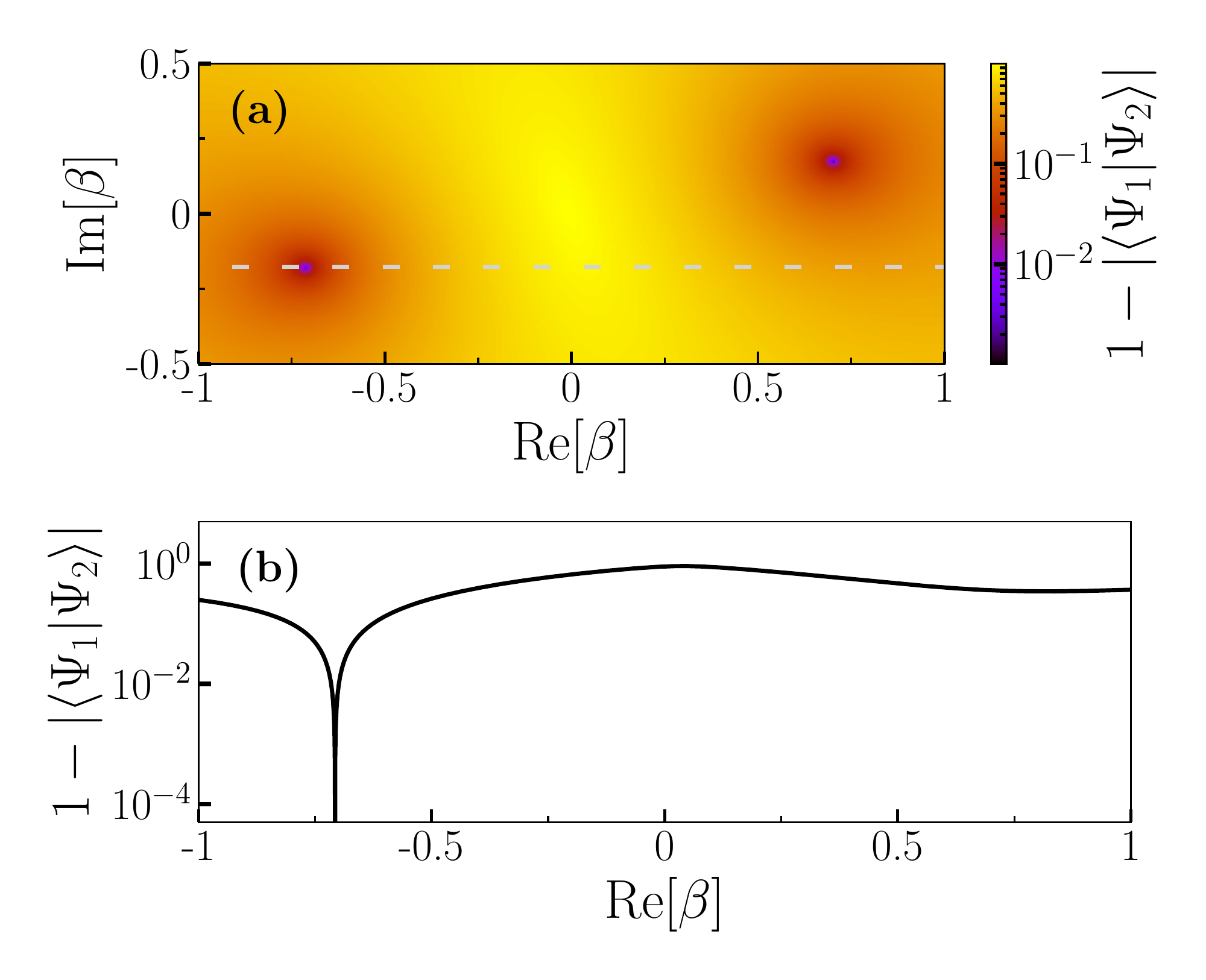}
    \caption{The overlap between the two eigenvectors $\ket{\Psi_{1,2}}$ of the NHH in \eqref{Eq:Vectors_1} as a function of $\beta$, having fixed $
    \gamma_+ = 0.5\gamma_-$ and $\omega = \gamma_-$.
    (a) The contourplot in the real and imaginary part of $\beta$ shows that, by appropriately choosing the value of $\beta$, it is possible to make $\ket{\Psi_{1,2}}$ coalesce. (b) Overlap along the gray-dashed line in panel (a), showing the abrupt change in the orthogonality properties of the eigenvectors near the EP.}
    \label{fig:Fig1}
\end{figure}

\begin{figure}[]
    \centering
    \includegraphics[width=0.49 \textwidth]{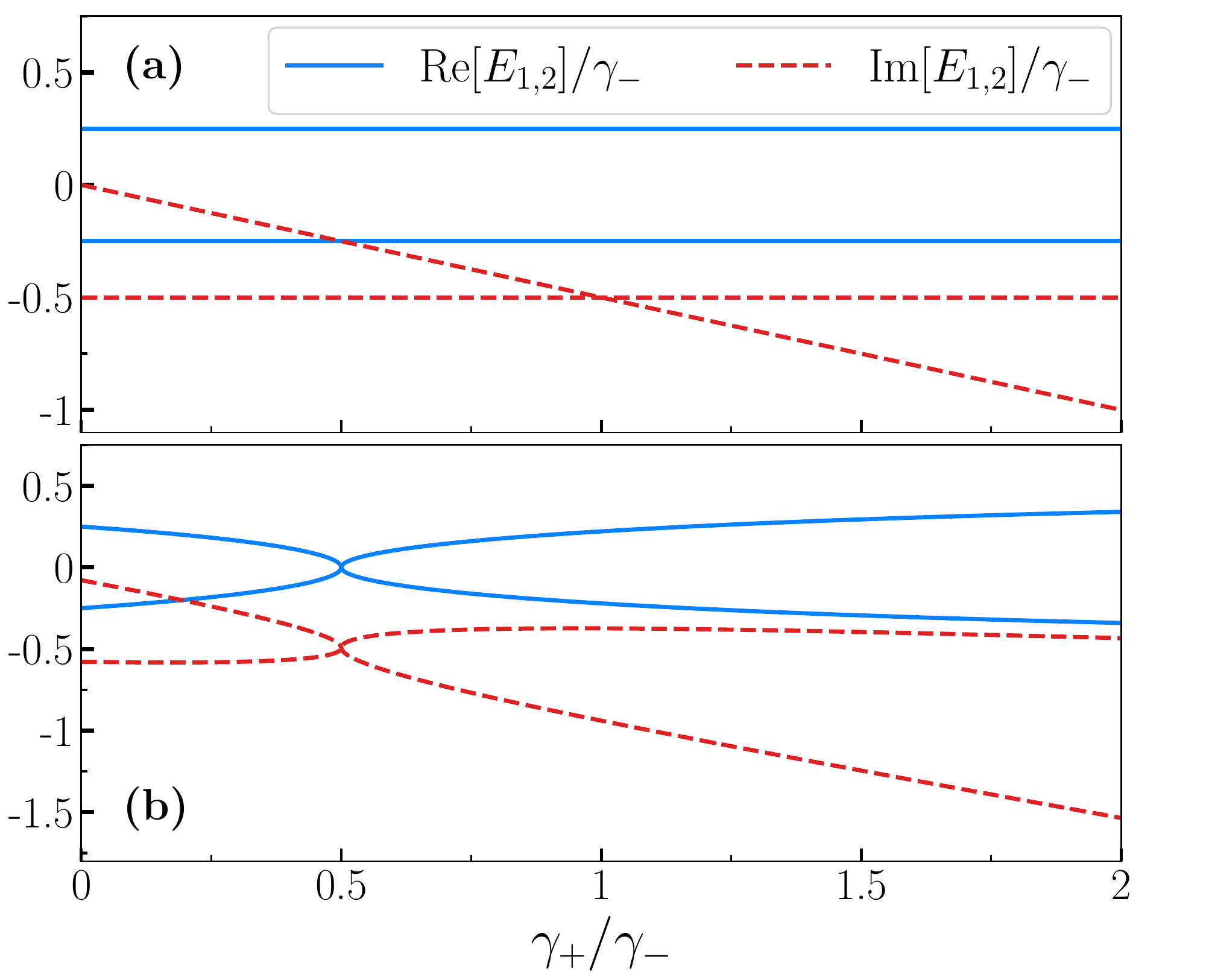}
    \caption{Real (blue solid curve) and imaginary (red dashed curve) parts of the two eigenvalues of the NHH in Eqs.~(\ref{Eq:Values_1})~and~(\ref{Eq:Vectors_1}) as a function of the gain over loss ratio $\gamma_+/\gamma_-$. (a) When $\beta=0$, the system does not exhibit an EP. (b) For nonzero $\beta\neq0$, an EP emerges the system [in this specific case $\beta=(2 + i)/4\sqrt{2}$]. We fix $\omega=\gamma_-$.}
    \label{fig:Fig2}
\end{figure}

\begin{figure}
    \centering
    \includegraphics[width=0.49 \textwidth]{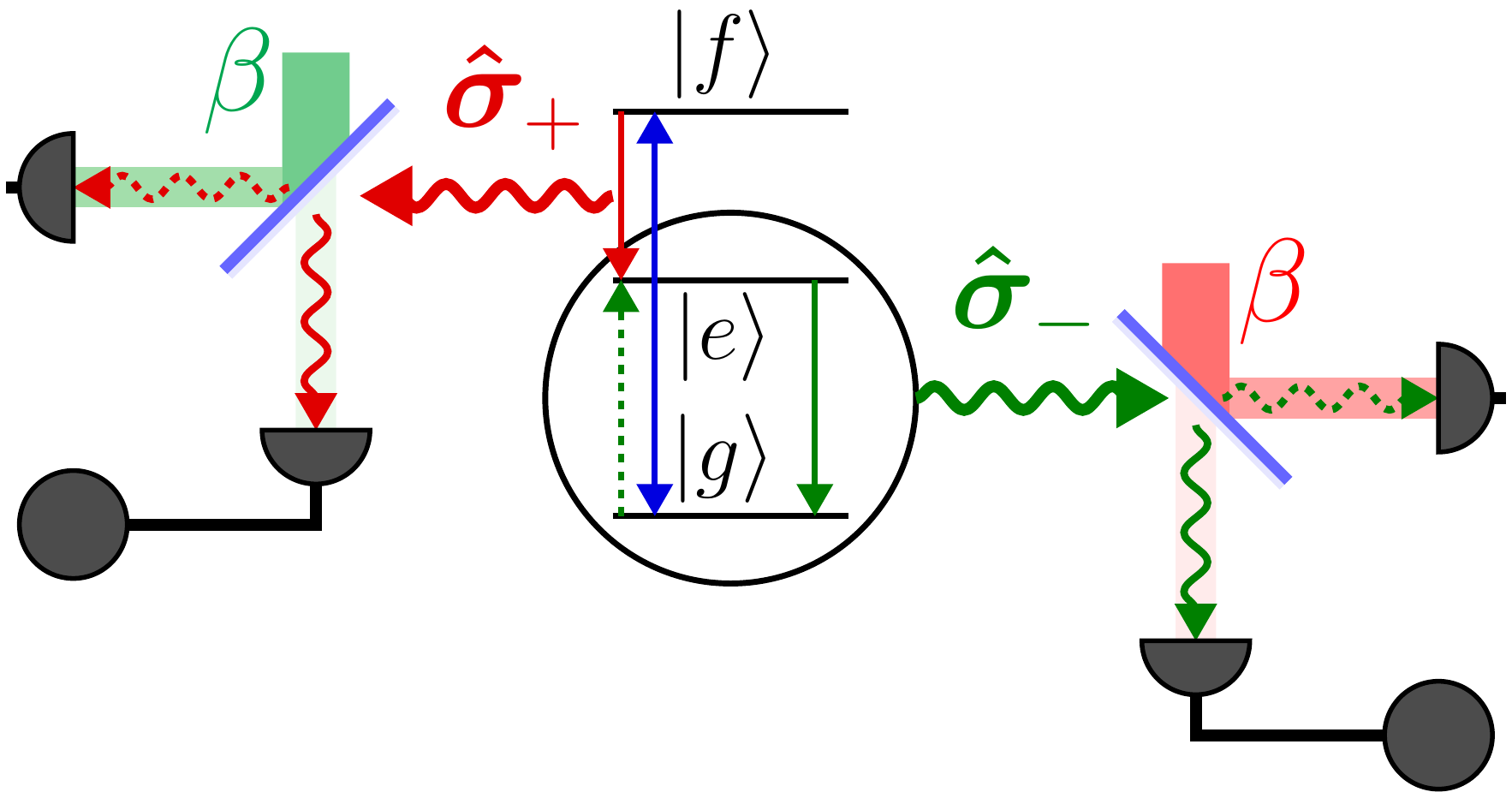}
    \caption{Setup to perform the $\beta$-dyne detection of a qubit with effective gain and saturation, leading to \eqref{Eq:Heff_beta_1}.
    A three-level system, characterized by the states $\ket{g}$, $\ket{e}$ (at energy $\omega$), and $\ket{f}$ (at energy $2\omega+\delta \omega$), is subject to decays $\ket{e}\to \ket{g}$ (green solid arrow) and $\ket{f}\to \ket{e}$ (red arrow) at rates $\gamma_{eg} \ll \gamma_{fe}$. A weak drive coherently couples the state $\ket{g}$ and $\ket{f}$ (blue arrow).
    The decay $\gamma_{eg}$, characterized by a jump operator $\hat{\sigma}_-$, is associated with the emission of an photon with energy $\omega$, and can be detected via the standard homodyne detection leading to \eqref{Eq:Lindblad_invariance_2}.
    The overall effect of the process $\ket{g}\to \ket{f} \to\ket{e}$
    is, instead, to induce an effective gain $\ket{g}\to\ket{e}$ (dashed green arrow) associated to the jump operator $\hat{\sigma}_+$, because of the rapid decay of $\ket{f}$.
    Such a gain, however, is accompanied by the emission of a photon with energy $\omega+\delta$, on which one can perform standard homodyne detection  and postselection.
    }
    \label{fig:Setup1}
\end{figure}

We show the effect of $\beta$ in Fig.~\ref{fig:Fig1}, where we plot the overlap between the two eigenvectors of $\hat{H}_{\rm eff} (\beta)$.
In particular, we notice that there is a whole region around the $\beta_{\rm EP}$ value where the eigenvectors almost coalesce, showing the dramatic effect that the introduction of $\beta$ can induce on the spectral properties of the NHH.
In Fig.~\ref{fig:Fig2}, instead, we show the emergence of the EP as a function of $\gamma_+/\gamma_-$, having fixed the value of $\beta$.
Compared to the $\beta=0$ case in Fig.~\ref{fig:Fig2}(a), we remark  that the eigenfrequencies now change both in real and imaginary parts as a function of $\gamma_+$.

\subsection{Physical realization}

Despite its relative algebraic simplicity, the previous examples require the simultaneous postselection of both the jumps occurring from $\hat{\sigma}^-$ and $\hat{\sigma}^+$.
While the former implies a spontaneous emission that can be, in principle, achieved via a high-fidelity detector, the latter corresponds to spontaneous excitation through gaining mechanisms, and its detection can be remarkably more difficult.
Such a proof-of-concept model can be realized, however, by using a three level system instead of a qubit \cite{Carvalho11,Santos11}.


{Consider a three-level system, whose undriven energy eigenstates $\ket{g}$, $\ket{e}$, and $\ket{f}$ are coupled by a weak coherent drive resonant with the transition between states $\ket{g}$ and $\ket{f}$ {(c.f. Fig.~\ref{fig:Setup1})}, according to the Hamiltonian
\begin{equation}\begin{split}
    \hat{H}(t) =& \ket{e}\bra{e} +  \left( 2 \omega + \delta \omega \right)  \ket{f}\bra{f}  \\ & \, + \Omega \left( \ket{g}\bra{f}e^{i\omega_{ef}t} + \ket{f}\bra{g}e^{-i\omega_{ef}t} \right).
    \end{split}
\end{equation}
The spontaneous emission of photons induces the decay of level $\ket{f}$ to $\ket{e}$, and from level $\ket{e}$ to the ground state $\ket{g}$, via the Lindbladian:
\begin{equation}
   {\cal L}_1= \gamma_{eg} \DD\left[\ket{g}\bra{e} \right]+ \gamma_{fe} \DD\left[\ket{e}\bra{f} \right].
\end{equation}
If we assume that $\gamma_{fe} \gg \gamma_{eg}, \Omega$, we can adiabatically eliminate the state $\ket{f}$. 
The combined action of the driving and the dissipation results in a new Lindbladian
\begin{equation}
    {\cal L}_2 = \gamma_{eg} \DD\left[\ket{g}\bra{e} \right]+ \gamma_\text{eff}\DD\left[\ket{e}\bra{g} \right],
\end{equation}
which therefore implements the wanted model with ${\gamma_+ \equiv \gamma_\text{eff}  = 4\Omega^2/\gamma_{fe}}$ and $\gamma_- \equiv \gamma_{eg}$. 
Notice that a jump from state $\ket{g}$ to $\ket{e}$ is associated with the emission of a photon at the frequency $(\omega+\delta\omega)$, which can thus be detected with a photon counter.
Such a jump can be distinguished from the one associated with $\ket{g} \to \ket{e}$,
because the latter leads to the emission of a photon at frequency $\omega$. }

\section{Example II: Inducing an EP in the driven Kerr resonator.}
\label{Sec:ExampleII}

\begin{figure}
    \centering
    \includegraphics[width=0.49 \textwidth]{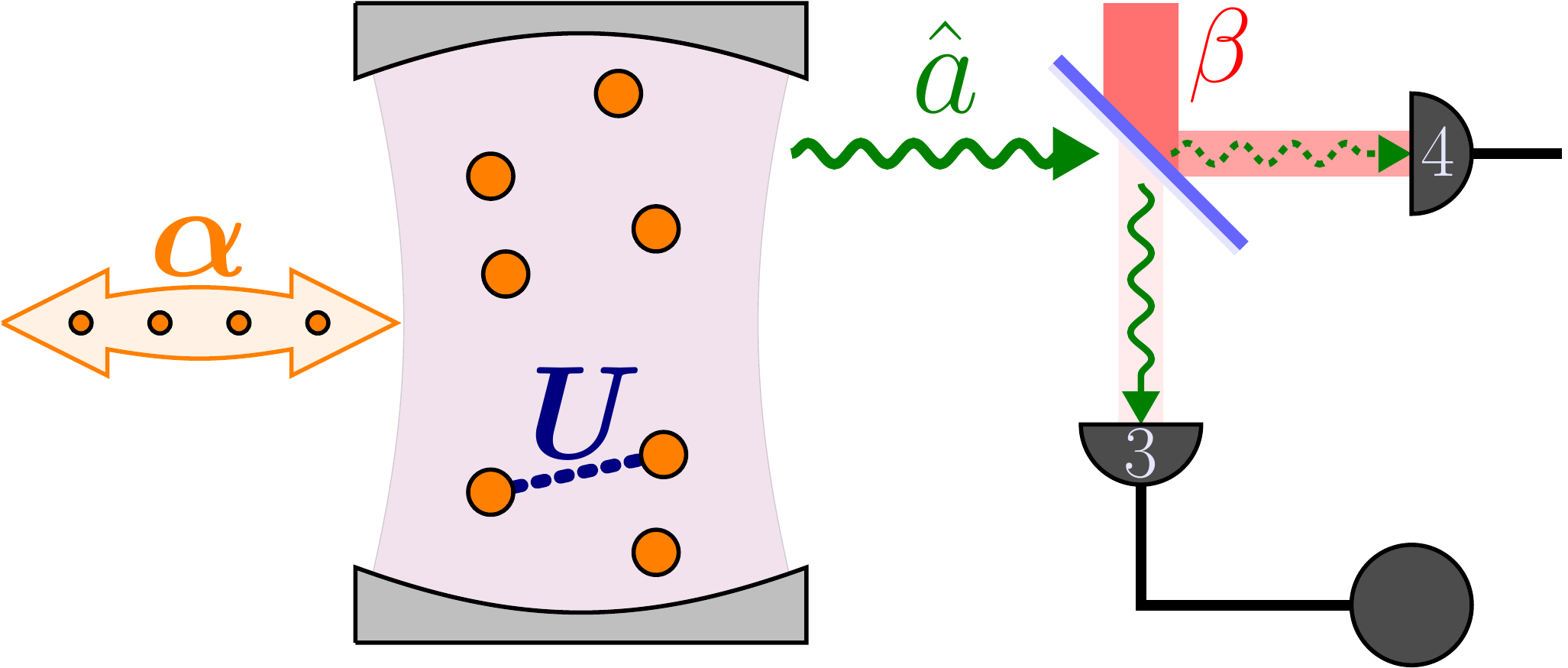}
    \caption{Setup to perform the $\beta$-dyne detection of a driven Kerr resonator, described by \eqref{Hkerr2}.
    The Kerr resonator is driven by a field of intensity $\alpha$, and two photons interact within the cavity with an intensity $U$.
    The emitted photons ($\hat{a}$ in green) are then mixed with a field whose effective intensity is $\beta$, as detailed in \eqref{Eq:Lindblad_invariance_2}.
    }
    \label{fig:Kerr_EP}
\end{figure}


In this section, we present a protocol based on a driven-dissipative Kerr resonator (see \figref{fig:Kerr_EP}), which requires detection of only emitted photons (i.e., the detection of one jump operator) in order to reveal the emergence of EPs in the postselected effective Hamiltonians.

The Hamiltonian of a Kerr resonator, in the frame rotating on the cavity frequency, is
\begin{equation}
    H = -\Delta \hat{a}^{\dagger} \hat{a} +  U\hat a^{\dagger2}\hat a^2 -i(\alpha \hat a^{\dagger}-\alpha^*\hat a),
\end{equation}
where $\hat a$ ($\hat a^{\dagger}$) is an annihilation (creation) field operator, $\Delta$ is the pump-to-cavity detuning, $U$ is a Kerr nonlinearity, $\alpha$ is the amplitude of a the coherent field which drives the resonator.
We assume that the system is subject to one-photon loss events described by the dissipator $\mathcal{D}[\hat{a}]$, occurring at a rate $\gamma$.

\begin{figure}[t!]
	\centering
	\includegraphics[width=0.5\textwidth]{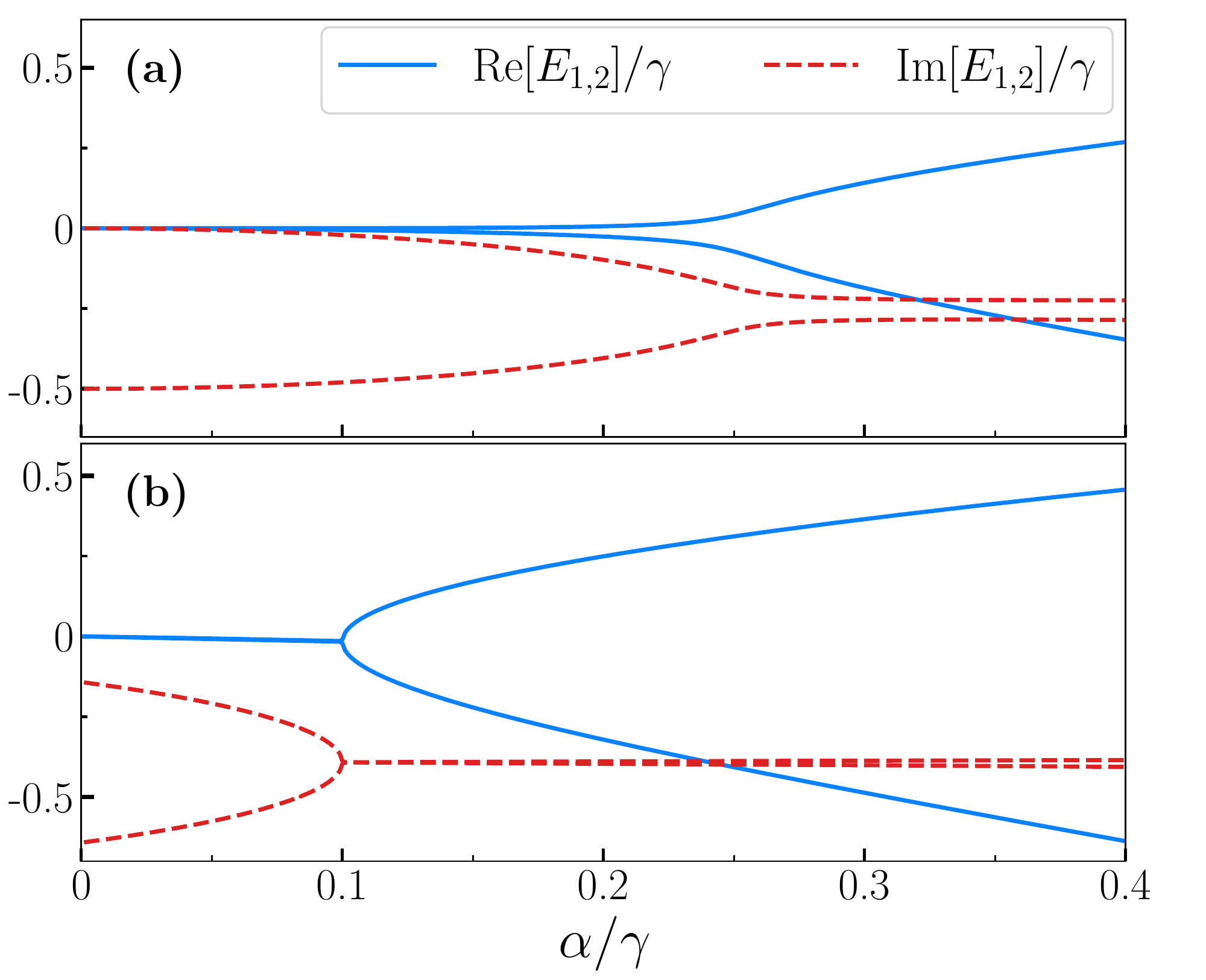}
	\caption{Real (blue solid curve) and imaginary (red dashed curve) parts of the two eigenvalues of the NHH in \eqref{Hkerr2} as a function of (purely) real-valued $\alpha$. (a) When $\beta=0$, the system does not exhibit an EP. (b) At certain nonzero $\beta\neq0$, an EP may be induced in the system (in this specific case $\beta=-0.5275-0.078i$). The rest of the system parameters are $\gamma=1$, $U=2$.}
	\label{Kerr_eig}
\end{figure}


Let us now assume that the system is at resonance, $\Delta=0$, and it is postselected in a $\beta$-dyne picure; thus, the NHHs $\hat H_{\rm eff}(\beta)$ is coherently displaced by the amplitude $\beta$, according to \eqref{Eq:NHH_beta}.
Assuming the weak-driving limit, i.e., $\alpha\ll\gamma , U$, and assuming also that $\beta\ll\gamma$, one can write down the effective Hamiltonian in the two-photon limit (i.e., truncating the Fock space at two photons), resulting in a three-level system that reads
\begin{equation}\label{Hkerr2}
    H_{\rm eff} = \begin{pmatrix}
    0 & i\alpha^* -i\gamma\beta^*& 0 \\
    -i\alpha & -\frac{i}{2}\gamma & i\sqrt{2}(\alpha^* -\gamma\beta^*)\\
    0 & -i\sqrt{2}\alpha & -i\gamma+2U
    \end{pmatrix}-\frac{i\gamma|\beta|^2}{2}{\mathbb I}_3.
\end{equation}
If $\beta=0$, no combination of parameters of the NHH results in an EP. 
In other words, whenever one is monitoring the system's environment and  leaked photons are not ``displaced'' by a coherent field, then the corresponding postselected NHH $\hat H_{\rm eff}(\beta=0)$ does not have any spectral singularity [see, e.g., \figref{Kerr_eig}(a)].

\begin{figure}
    \centering
    \includegraphics[width=0.5\textwidth]{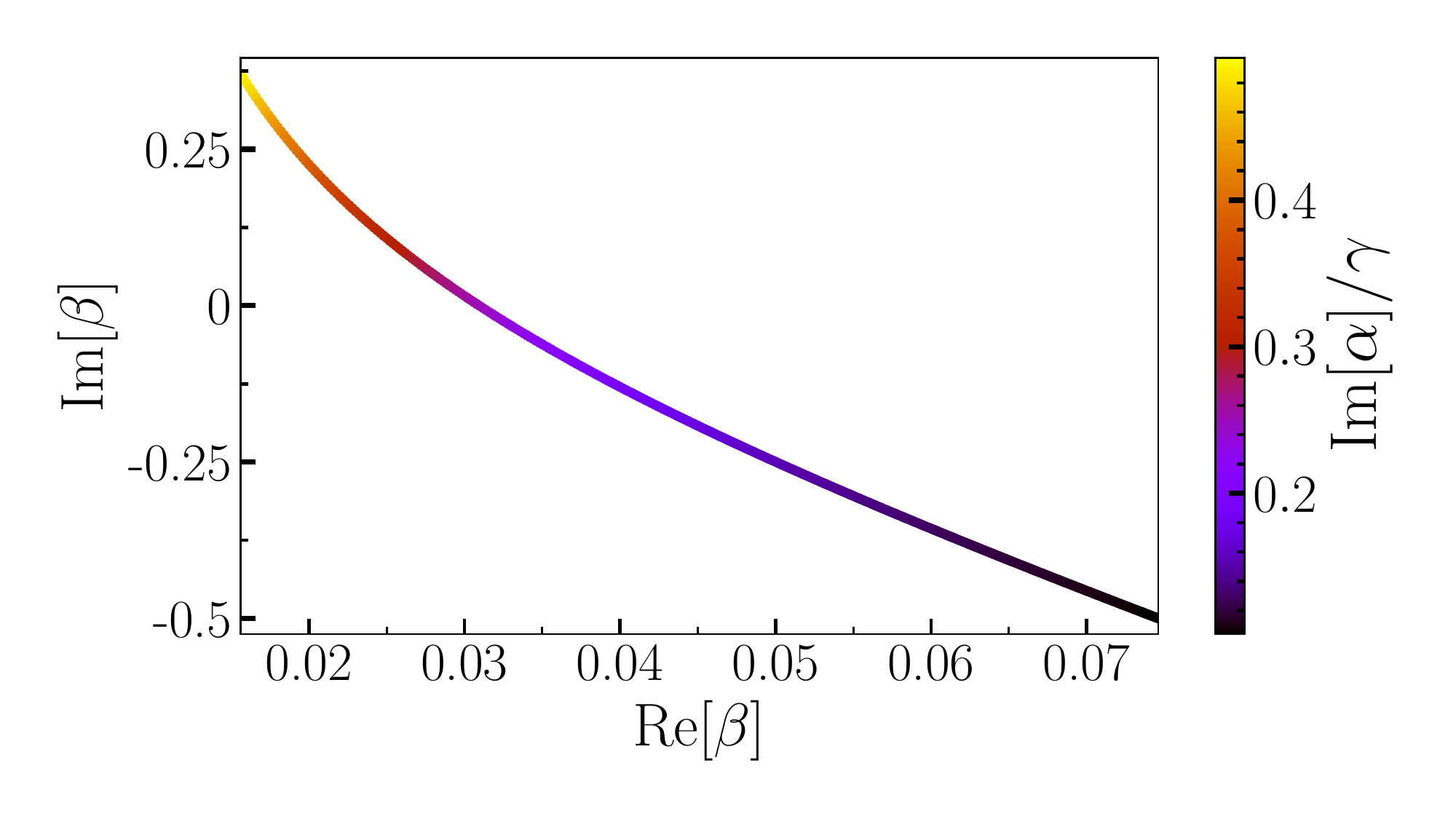}
    \caption{Values for the real and imaginary parts of the displaced field $\beta$, for a given purely imaginary-valued $\alpha$,  at which the NHH $\hat H_{\rm eff}(\beta)$ can exhibit an EP, according to \eqref{EPcond}. Other parameters are as in \figref{Kerr_eig}.}
    \label{Kerr_beta}
\end{figure}

For $\beta \neq 0$, instead, the condition for the NHH to have an EP of the second order reads:
\begin{eqnarray}\label{EPcond}
&a^2-3b-9ab+27c+2a^3\nonumber \\
& -3\sqrt{3}\left[27c^2+(4a^3-18ab)c-a^2b^2+4b^3\right]^{\frac{2}{3}}=0,
\end{eqnarray}
  where 
  \begin{eqnarray*}
  a &=& \frac{3i\gamma}{2}-2U, \,
  b = 3\alpha\beta^*\gamma-3|\alpha|^2-\frac{1}{2}\gamma^2-iU\gamma, \\
  c &=& (i\gamma-2U)\left(\gamma\beta^*\alpha-|\alpha|^2\right). 
  \end{eqnarray*}
Having fixed the system parameters, these equations can be numerically solved to find the values of $\beta$ resulting in an EP (see \figref{Kerr_beta}).
Note that only a second-order EP can be observed in the $\hat H_{\rm eff}(\beta)$, since at most only two of the three eigenvalues of the NHH  coincide.
 
We conclude that, if one \textit{postselects} those experiments where the detector \textit{never} clicks, and for an appropriate choice of $\beta$, satisfying the condition in \eqref{EPcond}, then the effective Hamiltonian in \eqref{Hkerr2} exhibits a second-order EP in the system [see \figref{Kerr_eig}(b)].
We stress that the full Lindbladian of the system does \textit{not} have a Liouvillian EP, in the sense described in Ref.~\cite{Minganti2019}. 
Indeed, the Linbladian describes the dynamics of the  operator averages and, as such, its averaging over various quantum trajectories eventually smears out EPs which emerge in certain effective NHHs. 
The presence of an EP in the NHH is thus an effect which can only  really emerge at the single-trajectory level.


%

\section{Example III: shifting an EP}
\label{Sec:ExampleIII}





The effect of $\beta$ is not only to generate new EPs in a quantum system; adjusting $\beta$ also allows one to \textit{shift} the position and \textit{change} the nature (i.e., the associated eigenvalues and eigenvectors) of an existing EP.

Consider a resonantly driven two level system, whose Hamiltonian in the pump frame reads
\begin{equation}
    \hat{H} = \frac{\omega}{2} \hat{\sigma}_x.
\end{equation}
Let us assume that the system is affected by a dissipation channel $\gamma_- \DD[\hat\sigma_-]$. 
The effective non-Hermitian Hamitonian of such a system in the standard representation is given by
\begin{equation}
\hat{H}_{\rm eff} = \frac{1}{2} \left(
\begin{array}{cc}
  -i \gamma_- & \omega \\
 \omega &  0 \\
\end{array}
\right).
\end{equation}
which exhibits an exceptional point for 
\begin{equation}\label{eq:excpoint_ex2}
    \omega = \frac{\gamma_-}{2}.
\end{equation}

We now assume that the emitted photons are detected via the $\beta$-dyne scheme. The associated non-Hermitian Hamiltonian is:
\begin{equation} \label{Eq:EP_betadyne}
\begin{split}
    &\hat{H}_{\rm eff}(\beta) = - i \frac{|\beta|^2 \gamma_-}{2} \hat{\mathds{1}}  \\ & \quad + \frac{1}{2} \left(
\begin{array}{cc}
- i\gamma_-  & \omega  \\
 \omega - 2 i\beta^*\gamma_- & 0  \\
\end{array}
\right),
\end{split}
\end{equation}
whose eigenvalues are
\begin{widetext}
\begin{equation}
    E_{1,2} = \frac{-i\gamma_- + 2i\vert\beta\vert^2\gamma_- \pm \sqrt{-\gamma_-^2 + 4\omega^2 + 8i\gamma_-\omega\beta^*}}{4},
\end{equation}
and the eigenvectors are
\begin{equation}
    \vert\Psi_{1,2}\rangle = \left\{-i\gamma_- \pm \sqrt{-\gamma_-^2 + 4\omega^2 + 8i\gamma_-\omega\beta^*} , {2\left(\omega + 2 i\beta^*\gamma_-\right)}\right\}.
\end{equation}
\end{widetext}
The exceptional point is now determined at
\begin{equation}
    \beta = i\frac{4\omega^2 - \gamma_-^2}{8\gamma_-\omega}.
\end{equation}
For $\beta = 0$ we retrieve \eqref{eq:excpoint_ex2}; for different values of $\beta$, the EP position, eigenvalues, and eigenvectors change.
We show this effect in Fig.~\ref{fig:Fig5}.
Notice that the shift of the EP comes at the expense of a decreased no-jump trajectory probability. A detailed computation of the postselection probability is reported in the Appendix \ref{App:BetaDyneModel}.

 \begin{figure}[h!]
     \centering
     \includegraphics[width=0.35 \textwidth]{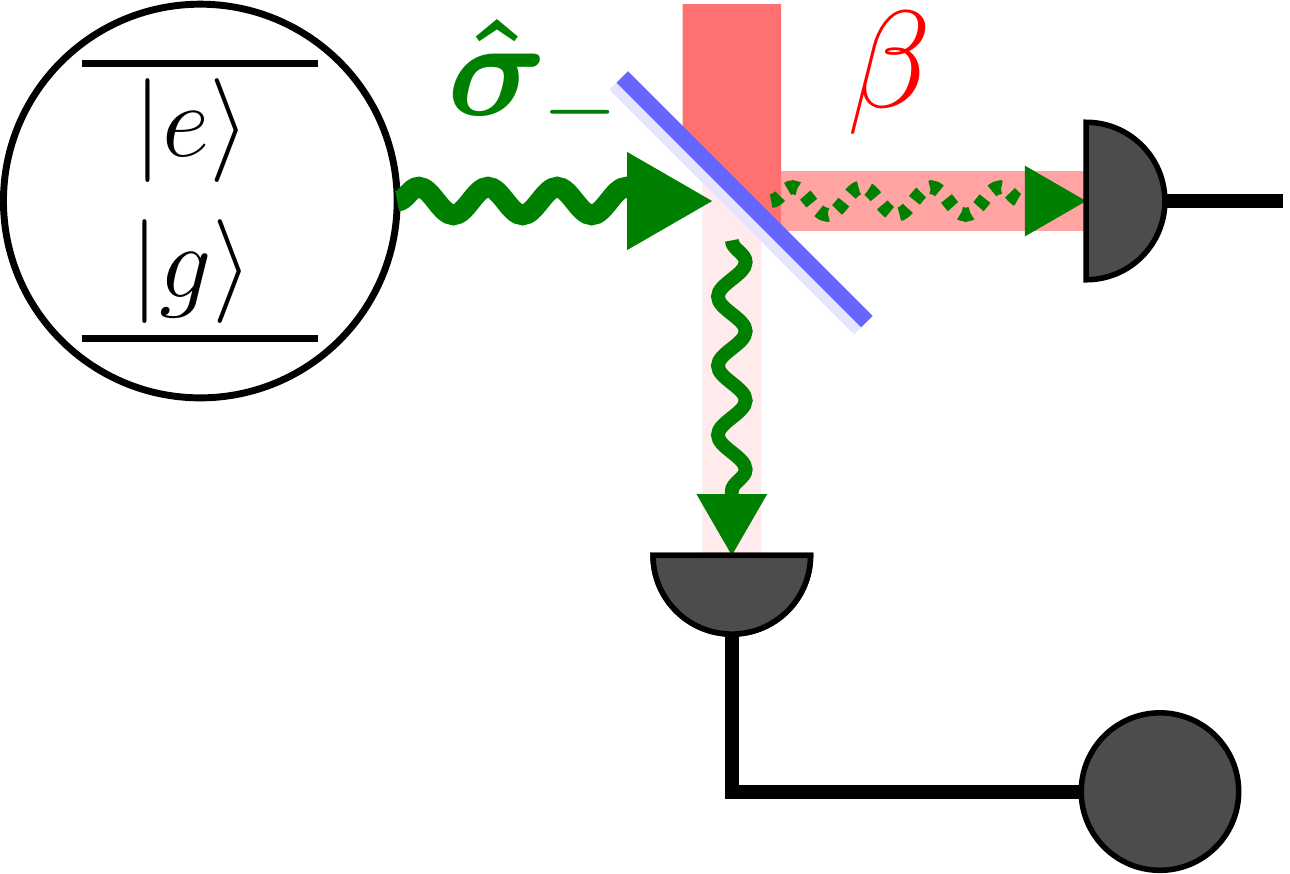}
     \caption{Setup to perform $\beta$-dyne detection of a driven qubit, described by \eqref{Eq:EP_betadyne}.
    The quibt is driven by a field  at resonance.
    The emitted photons are then mixed with a field whose effective intensity is $\beta$, as detailed in \eqref{Eq:Lindblad_invariance_2}.
    }
     \label{fig:Setup2}
 \end{figure}

\begin{figure}[t!]
    \centering
    \includegraphics[width=0.49 \textwidth]{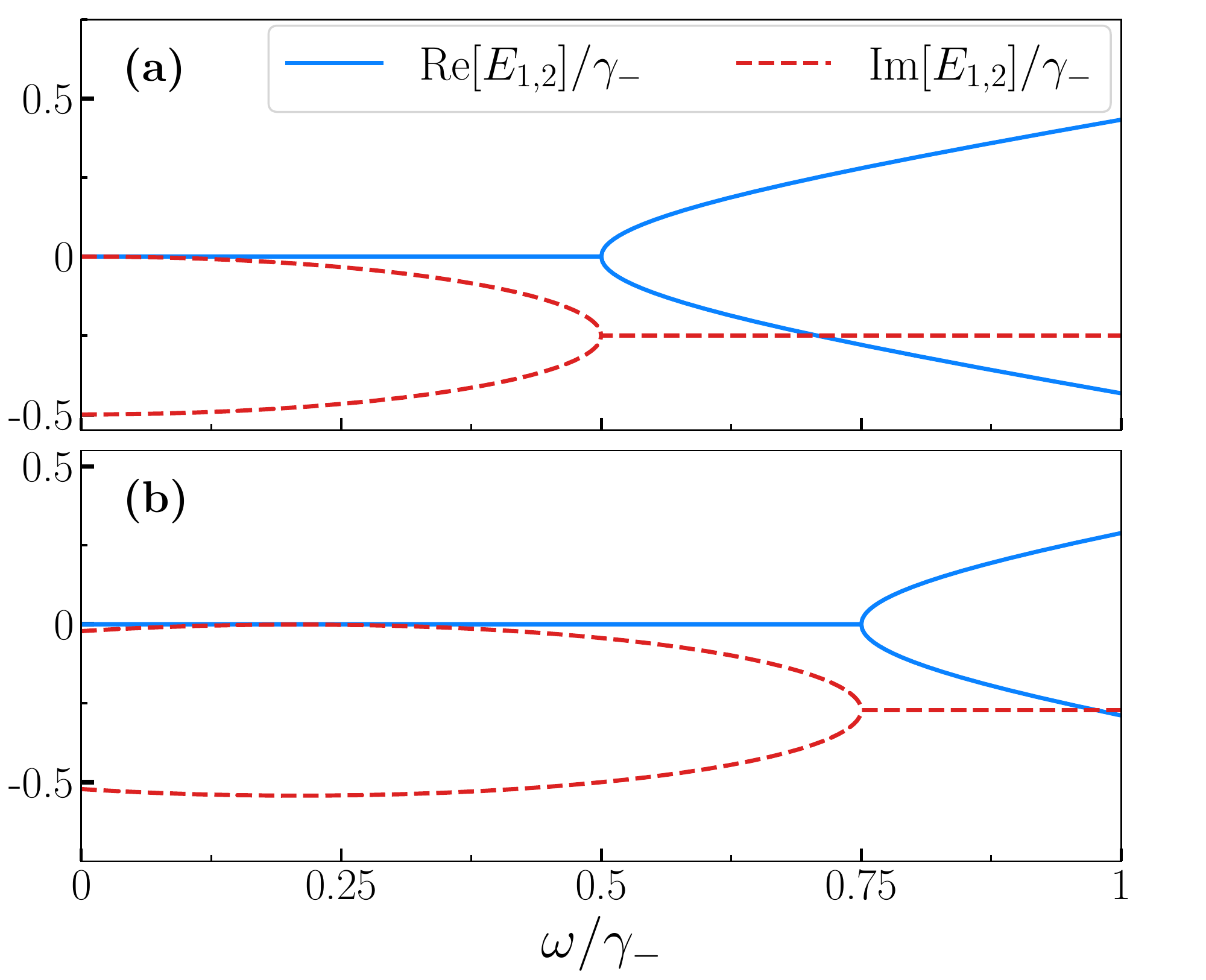}
    \caption{\textbf{(a)} $\beta = 0$ with an exceptional point at $\omega = \frac{1}{2}\gamma_-$ and \textbf{(b)} $\beta = \frac{5}{24}i$ with an exceptional point at $\omega = \frac{3}{4}\gamma_-$. }
    \label{fig:Fig5}
\end{figure}

A final comment is necessary.
As we already discussed, the postselection procedure of the Lindblad master equation can be, in principle, associated with different types of unraveling.
There are myriads of possible trajectories, including those with various values of $\beta$, inducing different coherent displacements [due to the Lindbladian invariance in \eqref{Eq:Lindblad_invariance_2}]. 
{This illustrates the fundamental difference between Liouvillian and Hamiltonian EPs. 
Liouvillian EPs exist at the level of the average dynamics of an open quantum system, and appear independently of the specific characteristics of the system-environment coupling. 
The EPs of a NHH, which are induced by post-selection, appear only for specific types of jump operators.
Furthermore, the fact that EPs of an NHH can be ``shifted'' by different unraveling via the action of $\beta$ is also a signature of the relative fragility of the EPs of postselected NHHs.
Indeed, both the system's parameters and those of the detector, whose action determines the effect of the postselection procedure, must be finely tuned.
}

\section{Conclusion}
\label{Sec:Conclusions}




By exploiting homodyne Lindbladian invariance, i.e., by displacing the emitted leaking photons with a laser field and keeping the whole Linbladian unchanged, we show that one can end up with different forms of quantum trajectories. Under postselection of no-jump trajectories, this scheme generates dynamics generated by a Non Hermitian Hamiltonian (NHH) whose spectral properties can be tuned via the parameters of the laser field used for the displacement. We illustrate the potential of this scheme on three examples based on simple quantum systems and realsistic detection setups, where Exceptionnal Points (EPs) can be generated or controlled only by changing the displacement of the leaking photons. This control on the EP however comes at the price of a more stringent postselection, decreasing the probability of occurrence of the no-jump trajectory.
More generally, our approach exploits the mathematical invariance of the Lindblad equation to provide a whole toolbox to engineer EP properties, opening the perspectives of new implementations of EP and their predicted applications, e.g. in metrology \cite{Hodaei17,Chen17} or optimal energy transfer \cite{Xu16,Assawaworrarit17,Khandelwal2021}.



\acknowledgments{
FM thanks the Laboratoire de physique in the ENS of Lyon, and Prof. Tommaso Roscilde, for the hospitality. DH acknowledges support from QuantERA ("MAQS" project). 
F.N. is supported in part by: Nippon Telegraph and Telephone Corporation (NTT) Research, the Japan Science and Technology Agency (JST) [via the Quantum Leap Flagship Program (Q-LEAP) program, and the Moonshot R\&D Grant Number JPMJMS2061], the Japan Society for the Promotion of Science (JSPS) [via the Grants-in- Aid for Scientific Research (KAKENHI) Grant No. JP20H00134], the Army Research Office (ARO) (Grant No. W911NF- 18-1-0358), the Asian Office of Aerospace Research and Development (AOARD) (via Grant No. FA2386-20-1- 4069), and the Foundational Questions Institute Fund (FQXi) via Grant No. FQXi-IAF19-06.
I.A. thanks the Project no.
CZ.02.1.01\/0.0\/0.0\/16\_019\/0000754 of the Ministry of
Education, Youth and Sports of the Czech Republic.}

\appendix

\section{Implementation of the $\beta$-dyne measurement setup and physical interpretation of Eq.~(\ref{Eq:Lindblad_invariance_2})}
\label{App:Homodyne}

The $\beta$-dyne measurement scheme is not only a mathematical object but can actually be implemented experimentally. It corresponds to a variation of the commonly used homodyne measurement protocol in quantum optics \cite{WisemanBook}.
In such setup, the signal (field emitted by the system) is mixed with an intense coherent field on an un-balanced beam-splitter with very low reflectance $\eta \ll 1$. The transmitted signal is then measured with a photon-counter (see also the scheme in Fig.~\ref{fig:Kerr_EP}).

The annihilation operators at the output ports $3,4$ of the beam-splitter are related to the input $1,2$ via:
\begin{align}
     \hat{a}_{3} &= \sqrt{1-\eta} \, \hat{a}_1+\sqrt{\eta}\, \hat{a}_2\\
     \hat{a}_{4} &= -\sqrt{\eta}\, \hat{a}_1+\sqrt{1-\eta}\, \hat{a}_2.\label{eq:BSrel}
\end{align}
For $\eta\ll 1$ and mode $\hat{a}_2$ prepared in a coherent state of amplitude $\vert\alpha_2\vert \gg 1$, we can see that the field at port $3$ is approximately described by 
\begin{equation}
\hat{a}_3 \simeq \hat{a}_1 + \sqrt{\eta}\alpha_2.
\end{equation}
Finally, as field $\hat{a}_1$ is populated by the system's emission, the detection of a click at port 3 therefore corresponds effectively to a jump of the system, captured by a coherently-displaced jump operator of the form $\hat{a} + \beta \hat{\mathds{1}}$. Consequently, everything happens as if the $\beta$-dyne measurement effectively generated a ``virtual'' field driving the system. 
This dynamics, including the effect of the virtual field, can be experimentally demonstrated; e.g., from multiple repetitions of the $\beta$-dyne measurement starting in the same state, followed by a projective measurement on the system.

Below, we derive more rigorously the backaction of such measurement on the system so as to deduce the postselection probability for the $\beta$-dyne no-jump trajectory.\\

\section{$\beta$-dyne measurement setup: theoretical model for the backaction and postselection probability}
\label{App:BetaDyneModel}


To demonstrate the setup able to implement the $\beta$-dyne measurement we discuss throughout the article, we now consider an emitter with Hamiltonian $\hat{H}_S$, whose light emission is collected into the port $1$ of a beam-splitter. 
Port $2$ corresponds to the incoming laser, which is modelled by a coherent initial state of amplitude $\alpha_2$.
The two fields in port $1$ and $2$ are mixed via beam-splitter transformation, resulting in the fields in ports $3$ and $4$ (see Fig.\ref{fig:Kerr_EP} in the case where the emitter is a cavity).

We first derive the evolution of the emitter's state associated with detecting a number of photons $n_3$ ($n_4$) photons at port $3$ ($4$) which is encoded in operator $M(n_3,n_4)$:
\begin{equation}
    M(n_3,n_4) = {}_4\bra{n_4}{}_3\bra{n_3} \hat{U}(\Delta t)\ket{0}_1\ket{\alpha_2}_2.
\end{equation}
Here $\ket{n}_i$ corresponds to the Fock state containing $n$ photons at port $i \in \llbracket 0,4\rrbracket$. On the other hand the ports $1,2$ are prepared in the vacuum state and a coherent state of amplitude $\alpha_2$ (the local oscillator), respectively. Finally, we have denoted $\hat{U}(\Delta t)$ the joint unitary evolution of the emitter and the mode $a_1$ which is given up to order ${\cal O}(\Delta t)$ by
\cite{Lewalle20}:
\begin{align}
    \hat{U}(\Delta t) &= \hat{\mathds{1}} -i\Delta t \hat{H}_\text{eff}(0) - \sqrt{\gamma_-\Delta t}\, (\hat{a}_1^\dagger\hat c-\hat{a}_1 \hat c^\dagger),
\end{align}
with $\hat{H}_\text{eff}(0) = \hat{H}_S -i\frac{\gamma_-\Delta t}{2}\hat c^\dagger \hat c$ the NHH in the case $\beta = 0$. We have introduced $\hat c$  as the emitter's lowering operator, e.g., $\hat{\sigma}_-$ for a qubit, or $\hat{a}$ for a cavity, as detailed in the examples in the main text.

We can now express $\hat{a}_1$ in terms of $\hat{a}_{3,4}$ using \eqref{eq:BSrel}. We also express the state $\ket{0}_1\ket{\alpha_2}_2$ in the basis of the output modes, yielding the tensor product of coherent states $\ket{\sqrt{\eta}\alpha_2}_3\ket{\sqrt{1-\eta}\alpha_2}_4$. We then obtain:
\begin{widetext}
\begin{align}
     \hat{M}(n_3,n_4) &=  {}_4\bra{n_4}{}_3\bra{n_3}  \left(\hat{\mathds{1}} - i\Delta t \hat{H}_\text{eff} - \sqrt{\gamma_-\Delta t}\, (\sqrt{1-\eta} \hat{a}_3^\dagger- \sqrt{\eta}\hat{a}_4^\dagger) \hat c\right) \ket{\sqrt\eta \alpha_2}_3\ket{\sqrt{1-\eta}\alpha_2}_4\nonumber\\
     &= \bra{n_3}\sqrt\eta \alpha_2 \rangle\bra{n_4}\sqrt{1-\eta}\alpha_2\rangle\left[\hat{\mathds{1}}- i\Delta t \hat{H}_\text{eff}(0)-\sqrt{\gamma_-\Delta t}\left(\sqrt{1-\eta}\frac{n_3}{\sqrt\eta\alpha_2}-\sqrt{\eta}\frac{n_4}{\sqrt{1-\eta}\alpha_2}\right)\hat{\sigma}_-\right].
\end{align}
\end{widetext}

We now assume that only the photons coming from port $3$ are detected. The state after a $\beta$-dyne measurement is therefore obtained by averaging over the values of $n_4$:
\begin{align}
    p(n_3)\hat{\rho}(t+\Delta t) & = \sum_{n_4}\hat{M}(n_3,n_4)\hat{\rho}(t)\hat{M}^\dagger(n_3,n_4),
    \end{align}
where $p(n_3)$ is the probability of reading  $n_3$ photons.

Finally we use
\begin{align}
    \sum_{n_4} \vert\langle n_4\ket{\sqrt{1-\eta}\alpha_2}\vert^2 \sqrt{\eta}n_4 = \sqrt{1-\eta} \vert\alpha_2\vert^2,\nonumber\\
    \sum_{n_4} \vert\langle n_4\ket{\sqrt{1-\eta}\alpha_2}\vert^2 \sqrt{\eta}n_4^2 = \sqrt{1-\eta} \vert\alpha_2\vert^2(\vert\alpha_2\vert^2+1),
\end{align}
to find:
\begin{widetext}
\begin{align}
    p(n_3)\hat{\rho}(t+\Delta t) &= \vert\bra{n_3}\sqrt\eta\alpha_2\rangle\vert^2\bigg[\hat{\rho}(t)-\Delta t  (i\hat{H}_\text{eff}(0)\hat{\rho}(t)+\text{H.c.}) 
  -\sqrt{\gamma_-\Delta t}\left(\frac{\sqrt\eta\alpha_2^*}{\sqrt{1-\eta}}-\frac{\sqrt{1-\eta}n_3}{\sqrt\eta\alpha_2}\right)\hat c\hat{\rho}(t) + \text{H.c.}\nonumber\\
    & +\gamma_-\Delta t \left(\frac{\eta(\vert\alpha_2\vert^2+1)}{1-\eta}+\frac{(1-\eta)n_3^2}{\eta\vert\alpha_2\vert^2}\right)\hat c\hat{\rho}(t)\hat c^\dagger.\bigg],
 \end{align} 
 \end{widetext}
    
    We now assume that $\eta$ is very small (the beam-splitter transmission is almost unity) and the coherent field is intense $\vert\alpha_2\vert\gg 1$, while keeping $\sqrt{\eta}\vert \alpha_2$ finite. Actually, we take $\sqrt\eta \alpha_2 \equiv \sqrt{\gamma_-\Delta t}\beta \ll 1$. We keep only terms up to order $\eta$ and $\gamma_-\Delta t$ and distinguish two measurement outcomes: $n_3 = 0$ and $n_3 >0$. The first one is associated with backaction:
    \begin{widetext}
    \begin{align}
    p(n_3=0)\hat{\rho}^{(0)}(t+\Delta t) &\simeq \hat{\rho}(t)+\bigg(-i\Delta t\left[\hat{H}_\text{eff}(0) -i\gamma_-\beta^*\hat c - i\frac{\vert\beta\vert^2\gamma_-}{2}\hat{\mathds{1}}\right]\hat{\rho}(t)+\text{H.c.}\bigg)\nonumber\\
    &= \hat{\rho}(t)-i\Delta t\left( \hat{H}_\text{eff}(\beta)\hat{\rho}(t)+\hat{\rho}(t)\hat{H}_\text{eff}^\dagger(\beta)\right),\label{Eq:evol0}
 \end{align} 
    \end{widetext}

    where
    \begin{align}
      \hat{H}_\text{eff}(\beta) = \hat{H}_\text{eff}(0)-i\frac{\gamma_-\vert\beta\vert^2}{2}\hat{\mathds{1}} - \gamma_-\beta^*\hat c  
    \end{align} 
    is the $\beta$-dyne effective Hamiltonian as defined in the main text. Note that we have used that $\vert\bra{0}\sqrt\eta\alpha_2\rangle\vert^2 \simeq 1-\gamma_-\Delta t \vert\beta\vert^2$. 
    
    On the other hand, the backaction associated with photon detection is obtained by summing up the terms with $n_3>0$:
    \begin{widetext}

    \begin{align}
    p(n_3>0)\hat{\rho}^{(1)}(t+\Delta t) &\simeq \gamma_-\Delta t \vert\beta\vert^2\hat{\rho}(t)-\bigg(i\Delta t\left[\hat{H}_\text{eff}(0) -i\gamma_-\beta^*\hat{\sigma}_- - i\frac{\vert\beta\vert^2\gamma_-}{2}\hat{\mathds{1}}\right]\hat{\rho}(t)+\text{H.c.}\bigg)\nonumber\\
    & \qquad +\frac{1}{\beta} \langle\langle n_3\rangle\rangle \hat c\hat{\rho}(t) + \text{H.c.}+ \frac{1}{\vert\beta\vert^2} \langle\langle n_3^2\rangle\rangle\hat c\hat{\rho}(t)\hat c^\dagger\nonumber\\
    &\simeq \gamma_-\Delta t (\hat c+\beta)\hat{\rho}(t)(\hat c+\beta^*).\label{Eq:evol1}
 \end{align}
    \end{widetext}
    We have introduced the average $\langle\langle \cdot \rangle\rangle$ in the distribution $P(n_3) = \vert\bra{n_3}\sqrt\eta\alpha_3\rangle\vert^2$ which verifies $\langle\langle n_3^2 \rangle\rangle \simeq \langle\langle n_3\rangle\rangle = \gamma_-\Delta t \vert \beta\vert^2 \ll 1$. This demonstrates that the quantum jumps associated with this detection setup are described by operator $\hat{J}_1 = \hat c + \beta \hat{\mathds{1}}$. \\
    
    Finally, the probability of obtaining outcome $n_3=0$ is encoded in the trace of the right-hand side of equality \eqref{Eq:evol0}:
    \begin{equation}
    \begin{split}
        p(n_3 = 0) = & 1-\gamma_-\Delta t \vert \beta \vert^2 \\ &-\Delta t\text{Tr}\left[i\hat{H}_\text{eff}(0)\hat{\rho}(t) + \text{H.c.}\right]
    \end{split}
    \end{equation}    
    for a single time-step $\Delta t$. More generally, one therefore expects that the probability of the trajectory with no photon detection is decreased by a factor $\exp(-\gamma_-\vert\beta\vert^2t)$ with respect to the case where $\beta = 0$ (direct detection of the emission).

\bibliography{main.bbl}
\end{document}